Superconducting Transition and Phase Diagram of Single Crystal $MgB_2$

U. Welp, G. Karapetrov, W. K. Kwok, G. W. Crabtree

Materials Science Division, Argonne National Laboratory, Argonne, IL 60439

Ch. Marcenat, L. Paulius*

Département de Recherche Fondamentale sur la Matière Condensée, Service de

Physique, Magnétisme et Superconductivité, CEA-Grenoble, 38054 Grenoble, France

T. Klein, J. Marcus

Laboratoire d'Etudes des Propriétés des Solides, CNRS, BP 166, 38042 Grenoble, France

K. H. P. Kim, C. U. Jung, H.-S. Lee, B. Kang, S.-I. Lee

NCRICS and Dept. of Physics, Pohang University of Science and Technology, Pohang

790-784, Republic of Korea

The superconducting phase diagram of $MgB_2$ was determined from magnetization, magneto-transport and the first single-crystal specific heat measurements. A zero-temperature in-plane coherence length of 8 nm is determined. The superconducting anisotropy increases from a value around 2 near $T_c$ to above 4.5 at 22 K. For H || c a pronounced peak effect in the critical current occurs at the upper critical field. Evidence for a surface superconducting state is presented for H || c, which might account for the wide spread in reported values of .

MgB$_2$ has emerged as a fascinating new superconducting material [1] which in addition to its surprisingly high value of T$_c$ displays a variety of unusual properties. Electronic structure calculations [2] indicate a highly anisotropic, complex Fermi surface consisting of two disconnected sections: a three dimensional tubular network of mostly boron $\pi$-states and two dimensional cylindrical sheets derived mostly from boron $\sigma$-states. Some of these features have been observed in recent de Haas-van Alphen experiments [3]. The appearance of multiple superconducting gaps was predicted [4], with a large gap residing on the 2D sheets and a small gap on the 3D network. Specific heat [5] and spectroscopic measurements [6,7] give evidence for this scenario. In addition, calculations within the anisotropic Eliashberg formalism [8] indicate a strongly anisotropic electron-phonon interaction. However, one of the basic parameters describing an anisotropic superconductor, the anisotropy coefficient $\gamma = H_{c2}^{ab}/H_{c2}^{c} = \xi_{ab}/\xi_{c}$, is not well established for MgB$_2$. Here, $H_{c2}^{ab}$, $H_{c2}^{c}$, $\xi_{ab}$ and $\xi_{c}$ are the in-plane and out-of-plane upper critical fields and Ginsburg-Landau coherence lengths, respectively. Reported values vary widely ranging from 1.1 to 6 depending on the measurement technique and on sample type, i.e., single crystals [10-15], oriented films [16], aligned crystallites [17], or powders [18,19]. Recent torque [14] and thermal conductivity [15] measurements on single crystals [14] as well as magnetization measurements on powders [19] indicate that the anisotropy coefficient is temperature dependent increasing strongly with decreasing temperature.

Here we present a detailed study of the superconducting phase diagram of MgB$_2$ combining magnetization, M(T), magneto-transport and the first single-crystal specific heat, C$_p$(T), measurements. The transport and magnetization data were taken on the same crystal. The upper critical fields for in- and out-of-plane fields were determined from M(T) and $C_p$(T) data. A coherence length of $\xi_{ab}$(0) = 8 nm is obtained. Transport data reveal a pronounced peak-effect in the critical current density at $H_{c2}^c$. For fields above $H_{c2}^c$ extending up to 1.66x$H_{c2}^c$ we observe strongly non-ohmic transport behavior which we attribute to surface superconductivity. Upward curvature in $H_{c2}^{ab}$(T) results in a temperature dependent anisotropy that increases from about 2 near $T_c$ to above 4.5 at 22 K. We note that the occurrence of surface superconductivity could account for the wide variation in reported values for the anisotropy constant.

The $MgB_2$ crystals were prepared by heat-treating a 1:1 mixture of Mg and B under high pressure conditions [9]. The crystals are well shaped with straight hexagonal facets and smooth faces (see picture in inset of Fig. 1b) with typical size of 50 μm. The magnetization was measured in a commercial SQUID magnetometer. The specific heat was measured in an ac-specific heat calorimeter [20] optimized to detect signals from minute crystals (of the order of 50 ng).

Fig. 1 shows M(T) measured on warming after field cooling the sample for (a) H || c and (b) H || ab, respectively. Breaks in the slope of the temperature dependence of the magnetization indicated by the vertical dotted lines are clearly seen and mark the onset of superconductivity. With increasing field, there is an essentially parallel shift of the superconducting transition to lower temperatures. This shift is much more pronounced for H || c (note the different temperature scales in panels (a) and (b)) indicating a strong superconducting anisotropy of $MgB_2$ as discussed below. Fig. 1c) shows the heat capacity of a second crystal from the same batch. In zero field a clear step in $C_p$(T) with a width of about 2 K is observed. With increasing field the step stays well defined and the

step height decreases as is expected. However, in contrast to polycrystalline samples [5], the transition width remains essentially constant. Using an entropy conserving construction for defining $T_c$ a phase boundary is obtained that agrees with that determined from M(T) as discussed below. Thus, the data shown in Fig. 1 represent the thermodynamic bulk transition of $MgB_2$ into the superconducting state.

Fig. 2 shows the resistive transitions in various fields along the c-axis. The sample is characterized by a resistivity of =1.6 µ cm at 40 K and a negligibly small normal state magneto-resistance. With increasing field the resistive transition moves to lower temperature and broadens significantly. Similar broadening has been observed in previous studies on single crystals [10-13]. However, here we show that the broadening is strongly current dependent. Non-ohmic behavior appears at the onset of the transition, labeled $T_{on}$. With increasing current a steep resistive drop emerges at a lower almost current independent temperature. At even higher currents a non-monotonic, hysteretic resistivity behavior arises that is reminiscent of the peak-effect. Peak-effects, that is, sharp maxima in the temperature and/or field dependence of the critical current and the corresponding suppression of the resistivity, have been observed just below $H_{c2}(T)$ in a variety of low pinning superconductors [21]. The peak-effect occurs also right below the melting transition separating the vortex lattice from the vortex liquid state in clean, untwinned $YBa_2Cu_3O_7$ crystals [22]. However, the vortex liquid state in high-$T_c$ superconductors is essentially ohmic in contrast to the $MgB_2$-data shown in Fig. 2. These results will be discussed further below in conjunction with the phase diagram. For H || ab (not shown here) the resistive transitions do not broaden with field in agreement with previous reports [10-13], and the peak-effect is largely suppressed.

Fig. 3 summarizes the transport behavior in the peak-effect region at 1.5 T ∥ c. The current-voltage (I-V) characteristics after field-cooling to 20 K display pronounced hysteretic behavior. On first increasing the current, a sharp onset of dissipation occurs near a critical current of 10 mA whereas for decreasing current zero-dissipation is approached near 5.5 mA. All subsequent current ramps and also the I-Vs taken after zero-field cooling follow this curve. These results are a manifestation of a current induced transition from a meta-stable high-$I_c$ vortex phase into a stable low-$I_c$ phase [23]. As the sample is field cooled through $T_c(H)$ a high-pinning vortex phase nucleates and stays in equilibrium until the peak-effect temperature is reached (see inset of Fig. 3). At lower temperatures this phase may survive as supercooled meta-stable state. The application of a strong enough current dislodges vortices from their pinned meta-stable configuration and triggers a transition into the stable low-pinning state which does not change on subsequent current ramps. In zero-field cooled measurements the initial vortex configuration is the result of flux-gradient driven motion of vortices across the sample and a low-pinning state analogous to the current-induced state is created.

The magnetic, calorimetric and transport data are summarized in the phase diagram shown in Fig. 4. For H ∥ c the onset of superconductivity determined from M(T) and $C_p(T)$ coincide with each other and with the location of the peak effect within the experimental uncertainty. We identify this line with the upper critical field for the c-axis, $H_{c2}^c(T)$. The observed peak-effect is the one occurring just below $H_{c2}^c$. A zero-temperature value of $H_{c2}^c(0) \approx 3.5$ T can be estimated which, using the WHH relation [24] $H_{c2}^c(0) = 0.7 \, \Phi_0/2\pi \xi_{ab}(0)^2$, yields the zero-temperature coherence length $\xi_{ab}(0) \approx 8$ nm. From the resistivity data in Fig. 2 and the Drude relation $l = 3/[\pi N(0) \, v_F \, e^2]$ an electron mean free path of $l \approx 98$ nm can be estimated using the density of states $N(0) = 0.7/$(eV unit cell) [5] and in-plane Fermi velocity $v_F = 4.8 \times 10^7$ cm/sec [2]. The BCS coherence length is given by $\xi_0 = 0.18 \hbar v_F/(k_B T_c) \approx 19$ nm indicating that our $MgB_2$ crystals are in the clean limit, in agreement with previous reports [10-13].

The onset of non-ohmic transport with decreasing temperature defines a line in the phase diagram lying a factor 1.66 above the $H_{c2}^c$-line. This suggests that the resistive onset is a manifestation of the onset of surface superconductivity [25] at $H_{c3}$ which for a flat surface in parallel magnetic field occurs at $1.7 \times H_{c2}$. The surface superconducting state supports a finite critical current which induces non-ohmic transport behavior, however, there is no diamagnetic signal in the magnetization [25]. For H || c the surface-superconducting currents are flowing on the vertical side faces of the plate-like crystals. Although resistive transitions as shown in Fig. 2 could in principle result from filamentary conduction along impurity phases the observation of a single sharp, current-independent superconducting transition in zero field indicates an intrinsic mechanism. Within the experimental resolution there is no feature in the magnetization and specific heat data that would indicate a second superconducting phase. In addition, nearly identical resistivity results were obtained on a second, smaller crystal. We also note that the limits of the resistive broadening reported earlier [11] encompass the same coefficient, 1.7. Indications for the development of a surface superconducting state have also been obtained from the comparison of electrical and thermal transport data [15]. Furthermore, in a recent study [26] on the low-current resistive transitions of $NbSe_2$, results closely resembling those in Fig. 2 have been obtained and interpreted as signature of surface superconductivity. In that study it was also observed that for reasons not entirely understood the surface effects are strongly suppressed for in-plane magnetic fields just as is the case of $MgB_2$ presented here. For H || ab the phase boundaries determined from the resistive and magnetic onsets coincide within the experimental uncertainty as shown in Fig. 4. The nature of the surfaces of $MgB_2$ and their effect on superconductivity has attracted recent interest since surface electronic states on Mg as well as on B terminated surfaces were observed in bandstructure calculations [27,28] as well as in angle resolved photoemission experiments [29]. Although their influence on suprconductivity is still controversial it has been suggested that superconductivity at the ab-surfaces is suppressed [27]. This might account for the strongly reduced surface superconductivity for H || ab. Reported -values determined from resistivity measurements (usually the resistive onset is identified with $H_{c2}$) on crystals [10-13] as well as on c-axis oriented films [16] are generally low, in the range of 2 to 3. In contrast, magnetic measurements on either powder samples [18,19] or on single crystals [14] as well as thermal conductivity measurements [15] yield -values around 4 to 6 at low temperatures. Since surface superconductivity does not contribute to the magnetization nor the thermal conductivity but does induce non-linear response in the resistivity a discrepancy between both determinations by a factor of order 1.7 might be expected which is actually in reasonable agreement with the spread of the reported values.

While the upper critical field for H || c follows a conventional temperature dependence for the ab-directions a pronounced upward curvature of $H_{c2}(T)$. As a result the superconducting anisotropy is temperature dependent as shown in the inset of Fig. 4. Similar results have recently been obtained from torque [14], magnetization on powder samples [19] and thermal conductivity [15] measurements. At high temperatures (i.e.

low fields) has a value between 1.5 and 2. At temperatures below 32 to 33 K (fields around 0.5 T) increases rapidly and reaches values above 4.5 near 22 K. An upward curvature of the $H_{c2}(T)$-line can arise in clean superconductors due to non-local effects as seen for example in borocarbides [30]. However, in those materials the upward curvature occurs in all crystal directions, and the out-of-plane anisotropy is essentially temperature independent. An alternative origin of the temperature dependent anisotropy could lie in the two-gap structure of $MgB_2$. Since the small 3D gap is readily suppressed in applied fields [5.6] $MgB_2$ behaves like a quasi 2D superconductor in sufficiently high parallel fields. Thus a steep $H_{c2}^{ab}(T)$-line can be expected. The cross-over between predominantly 3D to 2D behavior occurs around 0.5 T.

In conclusion, the superconducting phase diagram of $MgB_2$ has been determined using magnetization, magneto-transport and the first single-crystal caloric measurements. The in-plane coherence length is 8 nm corresponding to $H_{c2}^{c}(0)$ 3.5 T. The superconducting anisotropy increases with decreasing temperature from a value around 2 near $T_c$ to above 4.5 at 22 K. For H || c a pronounced peak effect in the critical current occurs at the upper critical field. Evidence for a surface superconducting state is presented for H || c which might account for the wide spread of reported values for the anisotropy coefficient.

This work was supported by the U.S. Department of Energy, BES, Materials Science under contract W-31-109-ENG-38, by the National Science Foundation under grant No. 0072880 and by the Ministry of Science and Technology of Korea through the Creative Research Initiative Program.

Figure captions

Fig.1  a) and b) Temperature dependence of the magnetic moment measured on warming after cooling in the indicated fields. Breaks in the slope of M(T) mark the onset of superconductivity as indicated by the dotted lines. The inset in b) shows a photo of the crystal. The smooth faces and the Au-contacts are seen. The irregular shaped features on the surface are glue residues. c) Temperature dependence of the heat capacity in several fields. The 3T data was used to subtract the background signal.

Fig.2  Resistive transition measured on cooling in various fields and with various current densities, 1 mA corresponds to a current density of 360 A/cm$^2$. The onset of non-ohmic behavior and peak-effect are indicated. At 1.5 T and 10 mA the hysteresis arising between cooling and warming is indicated by arrows.

Fig.3  I-V characteristics at 20 K and 1.5 T || c measured for increasing and decreasing current after field cooling (solid arrows) and after zero-field cooling (dashed arrows). The inset shows the temperature dependence of the critical current in 1.5 T || c.

Fig. 4  Superconducting phase diagram of $MgB_2$ as determined from the magnetization, specific heat and transport measurements.

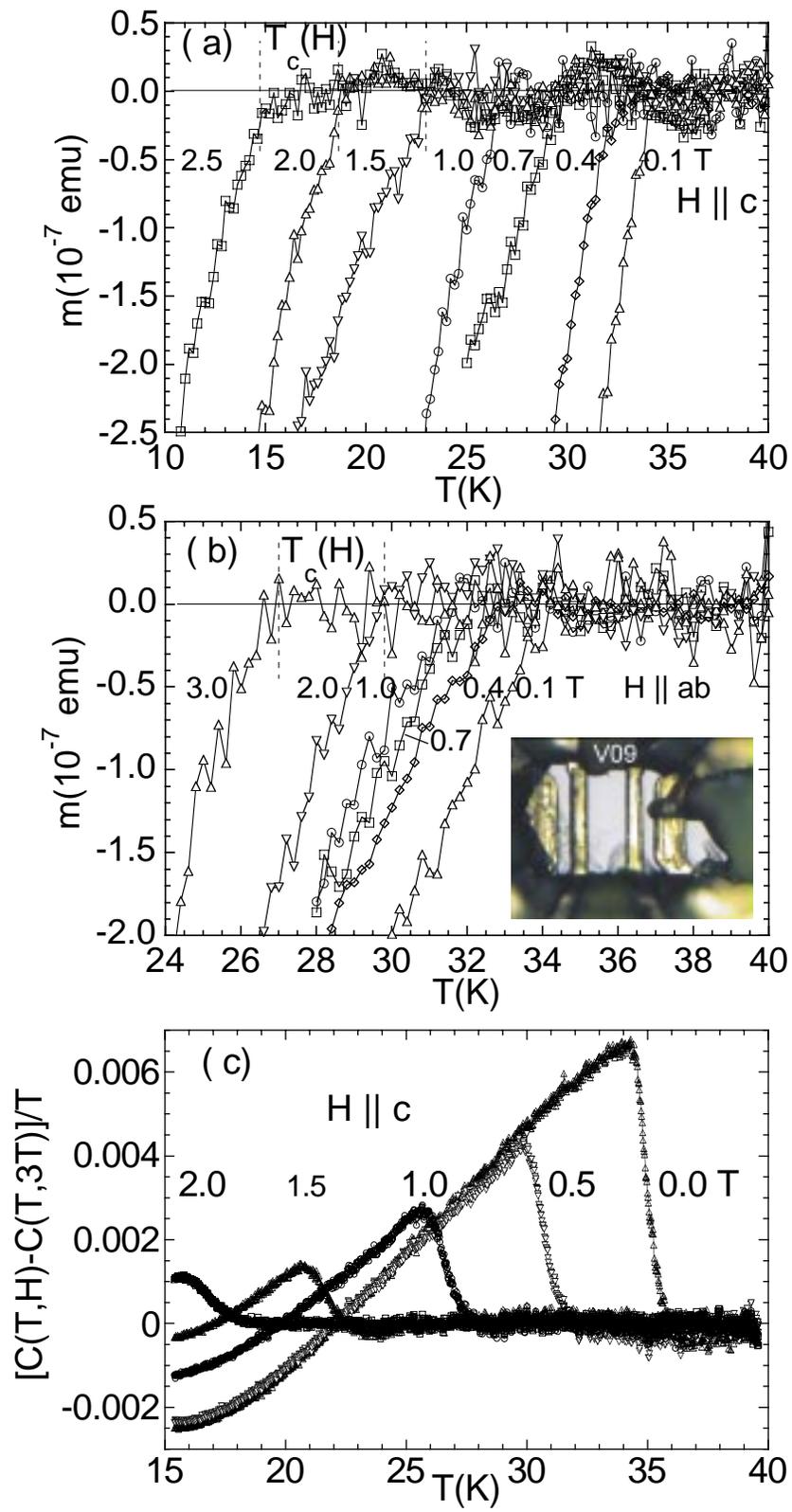

Fig.1
U. Welp et al.

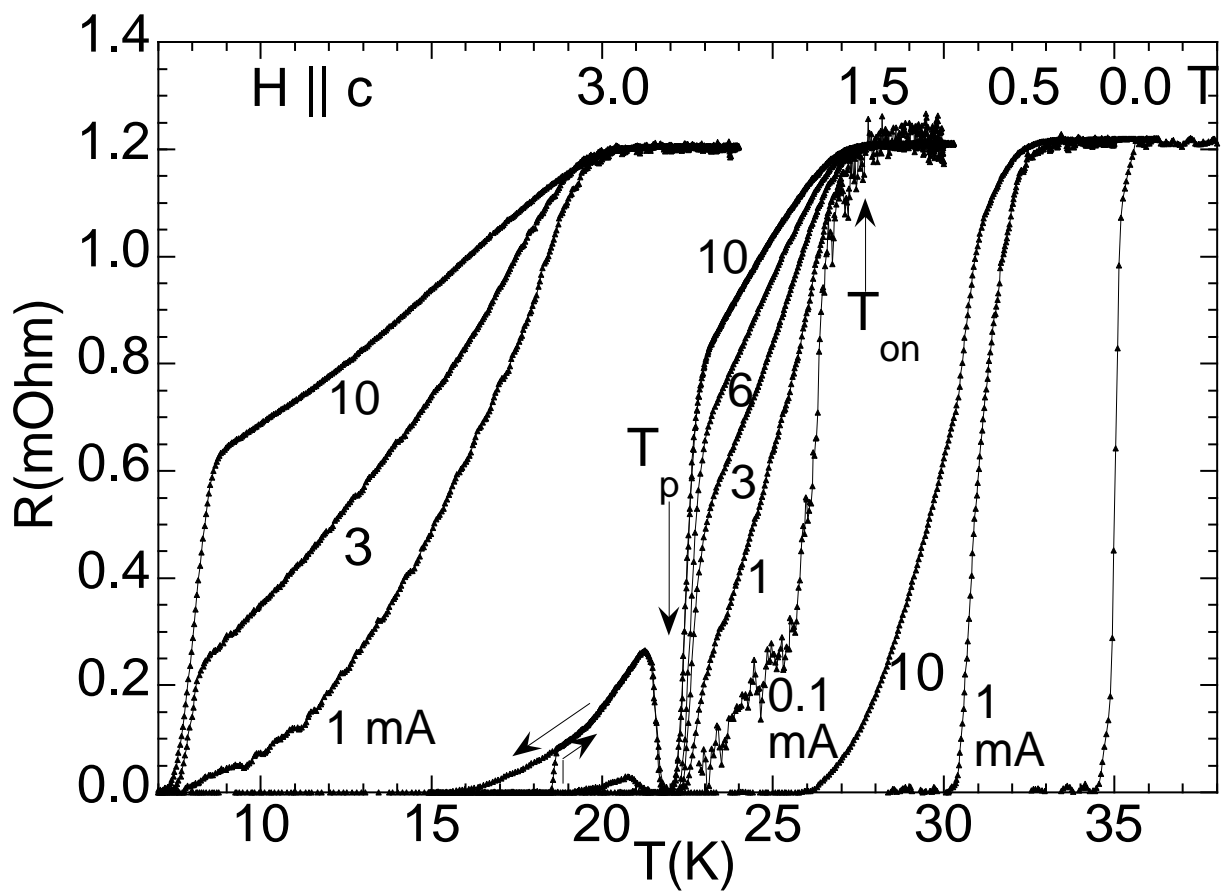

Fig. 2
U. Welp et al.

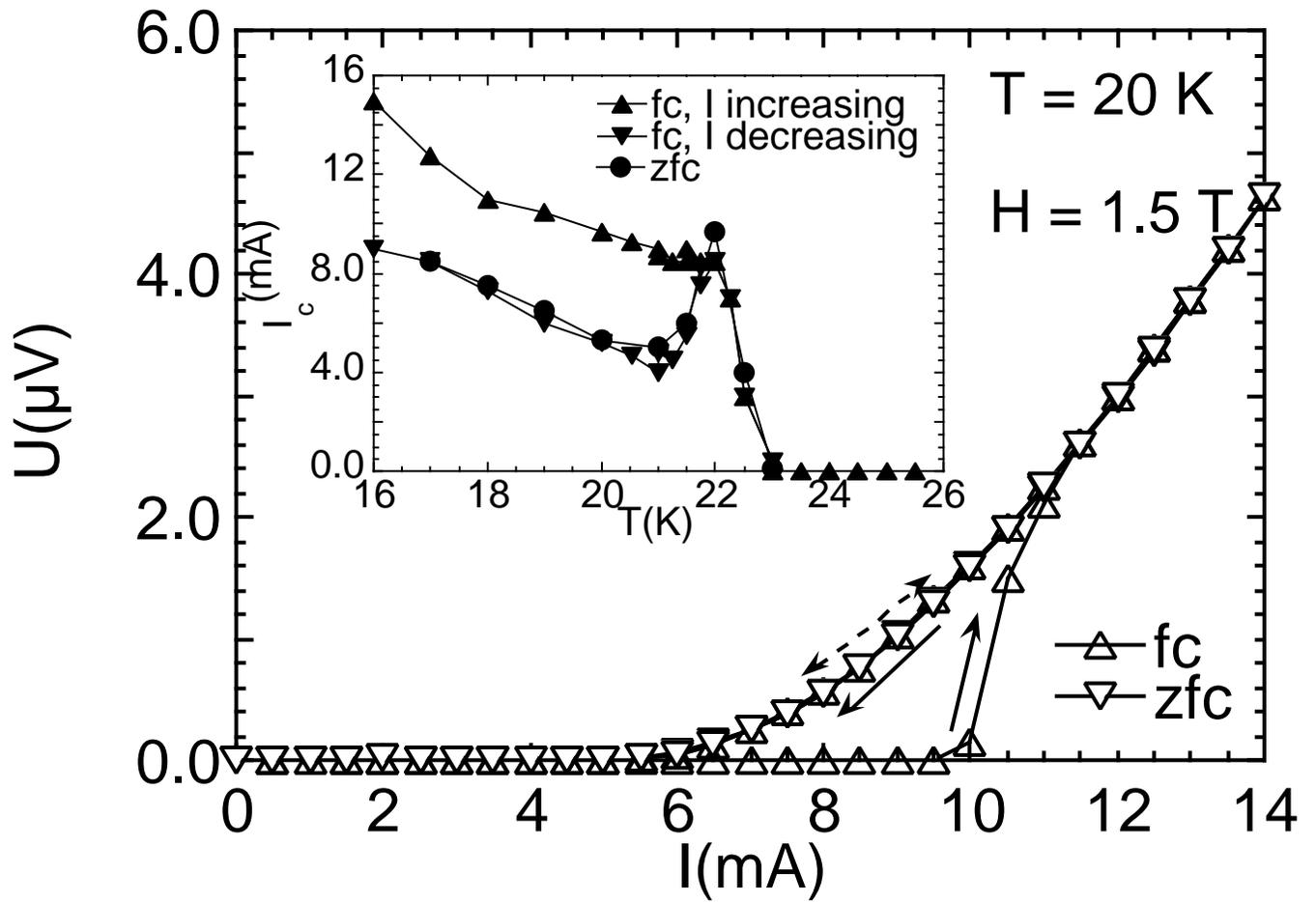

Fig. 3
U. Welp et al.

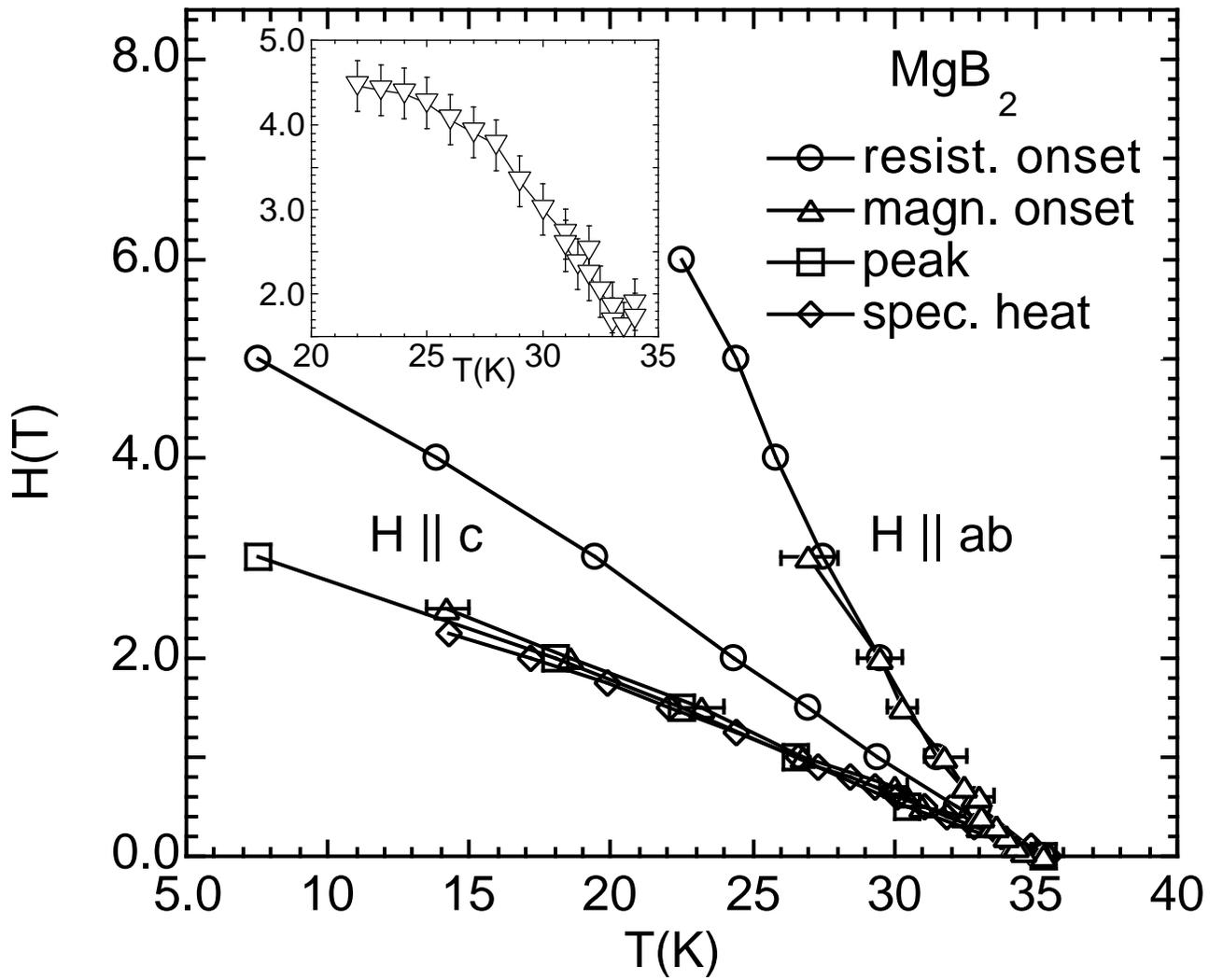

Fig. 4
U. Welp et al.